\documentstyle[12pt,aps]{revtex}

\input epsf

\begin{document}

\title{Chiral Symmetry: Pion-Nucleon Interactions in Constituent Quark Models}
\author{C.M.Maekawa and M.R. Robilotta }
\address{Instituto de F\'{\i}sica, Universidade de S\~ao Paulo \\
e-mail: maekawa@if.usp.br; robilotta@if.usp.br}

\date{\today}

\maketitle

We study the interactions of an elementary pion with a nucleon made of
constituent quarks and show that the enforcement of chiral symmetry requires
the use of a two-body operator, whose form does not depend on the
choice of the pion-quark coupling. The coordinate space
NN effective potential in the pion exchange channel
is given as a sum of terms
involving two gradients, that operate on both the usual Yukawa function
and the confining potential. We also consider an application to the case of
quarks bound by a harmonic potential and show that corrections due to the
symmetry are important.


\section{Introduction}

The pion nucleon ($\pi$N) form factor is present in a wide variety of
situations and  plays an important role in many hadronic processes. For
instance, in  elastic $ \pi$N scattering, it yields corrections to tree
diagrams  because the intermediate baryon states are off-shell. In the
case of  nucleon-nucleon interactions, on the other hand, it
corresponds  to an effective size that modifies the one
pion exchange  potential (OPEP) at short distances.

The $\pi$N form factor at intermediate energies is determined by the
cooperation of two complementary mechanisms, associated  with both
the meson cloud that surrounds the nucleon and with its
quark structure. In the former case, the relevant interactions may be
represented by 
diagrams with point-like nucleons,
such as those of fig.1, which were  recently
investigated in the framework of chiral perturbation  theory \cite
{ref1}.

The description of the intrinsic
extension of the nucleons, on the other hand,
requires models where quarks are bound by means of
bags\cite{ref2}, effective gluon interactions\cite{ref3}, non-relativistic
potentials or other mechanisms. Here, again, chiral symmetry is expected to play an
important role, but its implementation may prove to be more subtle.
Even if one starts from a chiral Lagrangian for free particles,
the
kinematical or dynamical approximations performed in the course of a
calculation involving bound systems
may effectively break the symmetry. This
situation is similar to the case of electromagnetic interactions with
deuterons or other nuclei, where 
gauge invariance is achieved with the help of
exchange currents. These currents are associated with binding effects,
since they arise from the coupling of the external photon with the fields 
that keep the nucleons together. 
Therefore, at least for non-relativistic
constituent quark models, one may expect the full implementation of
chiral 
symmetry also to require 
the inclusion of processes involving simultaneously
the pion probe and the binding fields.

The main purpose of this work is to study the role of chiral symmetry in
the pion-nucleon vertex, using a model in which constituent
quarks are confined by a generic scalar non-relativistic potential
and coupled to elementary  pions. 
We begin by deriving an effective NN potential, 
that is afterwards used to extract the $\pi$N form factor.

Our presentation is divided as follows. In sect.2 we introduce the basic
formalism and in sect.3  we present the effective chiral Lagrangians which
describe the interactions. The pion vertices are constructed in
sect.4, and the effective NN potential is obtained in sect.5. In sect.6
we apply our results to the case of a harmonic confining potential and
present conclusions in sect.7.


\section{Basic Formalism}

In this section we display our basic equations, 
with the purpose of establishing the
notation. The variables $\vec{r}_i$ refer to individual quarks, whereas $%
\vec{R}$, $\vec{\rho}$ and $\vec{\lambda}$ are collective and internal
coordinates of the nucleon, given by

\begin{eqnarray}
\vec{R} &=&\frac 13\left( \vec{r}_1+\vec{r}_2+\vec{r}_3\right) , \\
\vec{\rho} &=&\frac 1{\sqrt{2}}\left( \vec{r}_1-\vec{r}_2\right) , \\
\vec{\lambda} &=&\frac 1{\sqrt{6}}\left( \vec{r}_1+\vec{r}_2-2\vec{r}%
_3\right) .
\end{eqnarray}
\label{123}

\noindent
The centre of mass and relative coordinates of the two nucleon system are
denoted respectively  by $\vec{S}$ and $\vec{X}$ and related to the
individual coordinates $\vec{R}_{a}$ and $\vec{R}_{b}$ by

\begin{eqnarray}
\vec{S} &=&\frac 12\left( \vec{R}_a+\vec{R}_b\right) \;\;, \\
\vec{X} &=&\vec{R}_a-\vec{R}_b \;\;.
\end{eqnarray}

The effective Schr\"odinger equation for the two nucleon system is written as

\begin{equation}
\left\{ \frac{\nabla _S^2}{4M}+\frac{\nabla _X^2}M + \frac{P^{2}_{S}}{4M}
+E_X\right\} \left| N_a,N_b\right\rangle =V_{NN}\left( \vec{X}\right) \left|
N_a,N_b\right\rangle ,
\end{equation}
\label{6}

\noindent
where $P_S$ is the total momentum , 
$\left|N_a\right\rangle$ describes the collective motion of nucleon $a$ and 
$V_{NN}$ is the effective potential. For the six quark system, on the other
hand, we have

\begin{equation}
\left[ \sum_{i=1}^6\left( \frac{\nabla _i^2}{2m}\right) +\frac{P_S^2}{4M}%
+E\right] \left| q_1q_2q_3;q_4q_5q_6\right\rangle =V_{qq}\left(
q_1q_2q_3;q_4q_5q_6\right) \left| q_1q_2q_3;q_4q_5q_6\right\rangle .
\end{equation}
\label{7}

In this work we are interested in the pion-nucleon form factor, which is 
related to the potential at long and intermediate distances. Therefore we
assume that the six quark state may be decomposed into two clusters

\begin{equation}
\left| q_1q_2q_3;q_4q_5q_6\right\rangle =\left| q_1q_2q_3\right\rangle
\otimes \left| q_4q_5q_6\right\rangle \;\;.
\end{equation}
\label{8}

\noindent
This assumption corresponds to the idea
that there are no quark exchanges between the two nucleons. For the
quark-quark interaction, we write

\begin{equation}
V_{qq}=W+\Pi \;\;,
\end{equation}
\label{9}

\noindent
where $W$ is a short ranged confining potential and $\Pi $ is a long
ranged function due to the exchange of pions.

In agreement with the short-range nature of the confining potential and the
cluster decomposition of the six quark system, we assume that W operates
only inside each nucleon, and have

\begin{equation}
W\cong W_a+W_b \;\;,
\end{equation}
\label{10}

\noindent
where

\begin{equation}
W_i=W_i\left( \vec{\rho}_i,\ \vec{\lambda}_i\right) \;.
\end{equation}
\label{11}

Our last approximation consists in assuming that
pion exchanges are more relevant to interactions between quarks in different
nucleons than between quarks within a single cluster. Formally, this
corresponds to

\begin{equation}
\Pi \cong \Pi _{ab}=\sum_{v}\sum_{y}\Pi ^{\left(v,y\right)} ,
\end{equation}
\label{12}

\noindent
where the indices $v$ and $y$ refer to quarks in nucleons $a$ and $b$
respectively.

In order to display the quark structure of the nucleon, we write

\begin{equation}
\left| q_1q_2q_3\right\rangle =\left| N_a\left( \vec{R}_a\right)
\right\rangle \left| I_a\left( \vec{\rho}_a,\ \vec{\lambda}_a\right)
\right\rangle ,
\end{equation}
\label{13}

\noindent
where $\left| N_a\right\rangle$ and $\left| I_a\right\rangle$ correspond to
the collective and internal wave functions. The
latter is determined by the equation

\begin{equation}
\left[ - \frac{\nabla _{\rho _a}^2}{2m}-\frac{\nabla _{\lambda _a}^2}{2m}%
+W_a \right] \left| I_a\right\rangle =\epsilon _a\left| I_a\right\rangle .
\end{equation}
\label{14}

\noindent
Using this expression in eq.(7), we have

\begin{equation}
\left[ \frac{\nabla _S^2}{4M}+\frac{\nabla _X^2}M+\frac{\vec{P}_S^2}{4M}%
+E_X\right] \left| N_a,N_b\right\rangle \left| I_a\right\rangle \left|
I_b\right\rangle =\Pi _{ab}\left| N_a,N_b\right\rangle \left|
I_a\right\rangle \left| I_b\right\rangle,
\end{equation}
\label{15}

\noindent
with

\begin{equation}
E_X=E-\epsilon _a-\epsilon _b .
\end{equation}
\label{16}

Multiplying eq.(15) by $\langle I_b | \langle I_a |$, integrating over the
internal coordinates and comparing with eq.(6), we obtain the following
effective potential:

\begin{equation}
V_{NN}\left( \vec{X}\right) =\int d\Omega \left\langle I_b \right|
\left\langle I_a \right| \Pi _{ab} \left| I_a\right\rangle \left|
I_b\right\rangle \;\;,
\end{equation}
\label{17}

\noindent where

\begin{equation}
d\Omega =d\vec{\rho}_ad\vec{\lambda}_ad\vec{\rho}_bd\vec{\lambda}_b .
\end{equation}
\label{18}

The $\pi N$ form factor modifies the OPEP at short distances and may
be extracted from the effective potential. Denoting it by $G({k}^2)$, 
the modified OPEP may be written as

\begin{equation}
V_\pi \left(\vec{X},\left[ G\right] \right) =\left(\frac{g_{\pi NN}}{2M}%
\right)^2 {\cal{\vec{T}}}^{(a)}\cdot {\cal{\vec{T}}}^{(b)} \ \vec{%
\Sigma}^{(a)}\cdot \vec{\nabla}\ \vec{\Sigma}^{(b)}\cdot \vec{\nabla}\int 
\frac{d\vec{k}}{(2\pi)^3} \;\frac{e^{-i\vec{k}\cdot \vec{X}}}{{k}^2 + \mu^2}
G^2 ({k}^2) \;\;,
\end{equation}
\label{19}

\noindent
where $g_{\pi NN}$ is the $\pi$N coupling constant, $M$ and $\mu$ are the
nucleon and pion masses and $\vec{\Sigma}$ and ${\cal{\vec{T}}}$ are the
nucleon spin and isospin operators. Evaluating the gradients, we get

\begin{eqnarray}
V_\pi \left( \vec{X},\left[ G\right] \right) &=&\frac 13\left(\frac{g_{\pi
NN}\mu} {2M}\right)^2 \frac{\mu}{4\pi}{\cal{\vec{T}}}^{(a)} \cdot {%
\cal{\vec{T}}}^{(b)} \left\{ \vec{\Sigma}^{(a)}\cdot \vec{\Sigma}^{(b)}
\left[ U_0\left( X,\left[ G\right] \right) -D\left( X,\left[ G\right]
\right) \right] \right.  \nonumber \\
&+&\left.\left[ 3\vec{\Sigma}^{(a)}\cdot \hat{X}\ \vec{\Sigma}^{(b)} \cdot 
\hat{X}-\vec{\Sigma}^{(a)}\cdot \vec{\Sigma}^{(b)} \right] U_2\left(
X,\left[ G\right] \right) \right\} ,
\end{eqnarray}
\label{20}

\noindent where $D, U_0$ and $U_2$ are functionals of $G$, given by

\begin{eqnarray}
D\left( X,\left[ G\right] \right) &=&\frac 2{\pi \mu ^3}\int\limits_0^\infty
dk{k}^2G^2\left({k}^2\right) j_0\left( kX\right) \\
U_0\left( X,\left[ G\right] \right) &=&\frac 2{\pi \mu }\int\limits_0^\infty
dk\frac{{k}^2}{{k}^2+\mu ^2}G^2\left({k}^2\right) j_0\left( kX\right) \\
U_2\left( X,\left[ G\right] \right) &=&\frac 2{\pi \mu
^3}\int\limits_0^\infty dk\frac{{k}^4}{{k}^2+\mu ^2}G^2\left({k}^2\right)
j_2\left( kX\right)
\end{eqnarray}
\label{23}

The inversion of these results yields \cite{ref5}

\begin{eqnarray}
G^2\left({k}^2\right) &=&\mu ^3\int\limits_0^\infty dXX^2D\left( X,\left[
G\right] \right) j_0\left( kX\right) \;\;, \\
G^2\left({k}^2\right) &=&\frac{\mu ^{l+1}}{k^l}\left({k}^2+\mu ^2\right)
\int\limits_0^\infty dXX^2U_l\left( X,\left[ G\right] \right) j_l\left(
kX\right) \;\;.
\end{eqnarray}
\label{25}

In a more synthetic notation, we may also write

\begin{equation}
V_\pi \left( \vec{X},\left[ G\right] \right) = {\cal{\vec{T}}}^a\cdot {%
\cal{\vec{T}}}^b\left[ -\vec{\Sigma}^a\cdot \vec{\Sigma}^b \; V_0^{-}\left( 
\vec{X},\left[ G\right] \right) +S_{12} \; V_2^{-}\left( \vec{X},\left[ G\right]
\right) \right] \;,
\end{equation}
\label{26}

\noindent where

\begin{equation}
V_l^{-}\left( \vec{X},\left[ G\right] \right) =\frac 1{6\pi ^2\mu ^2}\left(
\frac {g_{\pi NN}\mu}{2M}\right) ^2 \int\limits_0^\infty dk\frac{{k}^4}{{k}%
^2+\mu ^2}G^2\left({k}^2\right) j_l\left( kX\right) \;\;.
\end{equation}
\label{27}

In this case, the form factor is given by 

\begin{equation}
G^2\left({k}^2\right) ={12\pi}{\mu}^2{\left( \frac{2M}{g_{\pi NN} \mu }%
\right)}^2 \frac{\left({k}^2+\mu ^2\right) }{{k}^2}\int\limits_0^\infty
dXX^2V_l^{-}\left( X,\left[ G\right] \right) j_l\left( kX\right) .
\end{equation}
\label{28}


\section{Dynamics}

We assume that the scalar confining potential W is associated with a field
S, that may represent effectively
the self interactions of gluons as, for instance, in the case of
non-topological solitons \cite{ref6,ref7,ref8,ref3}. Alternatively, it may
be associated with fluctuations of the chiral scalar field $\sigma ^{\prime }
$, considered long ago by Weinberg\cite{ref9}. The important point to stress,
however, is that our calculation is completely independent of the
meaning attached to this scalar field. For the pion sector we adopt the
non-linear sigma model and the basic Lagrangian is written as

\begin{equation}
{\cal{L }}=\left[ \frac 12\left( \partial _\mu S\partial ^\mu
S-m_s^2S^2\right) -V\left( S\right) \right] +\left[ \frac 12\left( \partial
_\mu \vec{\phi}\partial ^\mu \vec{\phi}+\partial _\mu \sigma \partial ^\mu
\sigma \right) +f_\pi \mu ^2\sigma \right] + {\cal{L}}_q ,
\end{equation}
\label{29}

\noindent
where $\vec \phi $ and $f_\pi $ are the pion field and decay constant,
whereas $\sigma $ corresponds to the function $\sigma =\sqrt{f_\pi ^2-\phi ^2%
}$. Formally, $V(S)$ represents a potential associated
with self interactions of the scalar field, but it has no direct role here.
The Lagrangian ${\cal{L}}_q$ represents both the quark sector
and its interactions with the bosonic fields.

There are many possible forms for the Lagrangian ${\cal{L}}_q$, two of
which are widely employed in the literature. In one of them the pion-fermion
coupling is pseudo-vector (PV), whereas in the other it is pseudo scalar
(PS). In the case of PV coupling, one has

\begin{equation}
{\cal{L}}_q^{PV} = \bar{\psi}i\gamma _\mu D^\mu \psi - m\bar{\psi}\psi + 
\frac{g}{2m}\bar{\psi}\gamma_\mu \gamma_5 \vec{\tau}\psi \cdot D^\mu \vec{\phi}
-g_{s} S\bar{\psi}\psi \;\;,
\end{equation}
\label{30}

\noindent
where $\psi$ and $m$ are the constituent quark field and mass, $g$ is the
pion-quark coupling constant and $g_s$ represents the coupling of the quark
to the scalar. In this expression the pion and nucleon covariant derivatives
are given by\cite{ref10}

\begin{eqnarray}
D^\mu \vec{\phi} &=&\partial _\mu \vec{\phi}-\frac 1{\sigma +f_\pi }\partial
^\mu \sigma \vec{\phi}, \\
D^\mu \psi &=&\left[ \partial ^\mu +i\frac 1 {f_\pi \left( \sigma +f_\pi
\right)}\frac{\vec{\tau}} 2\left( \vec{\phi}\times \partial ^\mu \vec{\phi}%
\right) \right] \psi .
\end{eqnarray}
\label{32}

\noindent For $PS$ coupling, on the other hand, the chiral Lagrangian for
the fermion sector is

\begin{equation}
{\cal{L}}_q^{PS}=\bar{q}i\rlap/\partial q-g\bar{q}\left( \sigma +i\vec{\tau%
}\cdot \vec{\phi}\gamma _5\right) q -{\left(\frac{g_s}{f_{\pi}}\right)}S\bar{%
q}\left( \sigma +i\vec{\tau} \cdot \vec{\phi}\gamma _5\right) q ,
\end{equation}
\label{33}

\noindent
where q is a quark field that transforms linearly.

On general grounds one knows that, in the framework of chiral symmetry,
results should not depend on the choice of ${\cal{L}}_q$\cite{ref11,ref12}.
However, this point is not always appreciated in particular calculations
and we would like to stress it in this problem. Therefore, we adopt both
forms of $ {\cal{L}}_{q} $ 
and demonstrate explicitly, in the next section,  that our results do
not depend on how the symmetry is implemented.


\section{Pion Vertices}

In this section we evaluate the operators needed to construct the effective
NN interaction. In order to motivate our approach we recall that, in 
general, chiral symmetry is realized by means of families of diagrams
organized according to loop and momentum counting rules. For instance, when
two free point-like nucleons interact through pion fields,
the simplest chiral family involves just a
single diagram, associated with the one pion exchange potential (OPEP)\cite
{ref13}. In the case of PV coupling, the $\pi N$ vertex used to construct
the OPEP is proportional to the matrix element

\begin{equation}
{\Gamma}_{\pi N}=\frac g{2m}\bar{u}\left(\vec p ^{\prime }\right) \rlap/%
k\gamma _5\vec{\tau}u\left(\vec p\right) ,
\end{equation}
\label{34}

\noindent
where $k = p^{\prime} - p$. Using the equations of motion for the nucleons,
we may rewrite ${\Gamma}_{\pi N}$ as

\begin{equation}
{\Gamma}_{\pi N}=g\bar{u}\left(\vec p^{\prime }\right) \gamma _5 \vec{\tau}%
u\left(\vec p\right) ,
\end{equation}
\label{35}

\noindent
which is the expression one would obtain from the PS Lagrangian. Hence
both couplings yield the same result. This kind of equivalence,
which must hold for all chiral families of processes, is true for the OPEP
only if the nucleon wave functions are exact solutions of the equation of
motion. When this does not happen, the PV and PS couplings do not yield the
same results, indicating that the single pion exchange no longer
constitutes an autonomous chiral family.

In the case of composite nucleons, this result means that single pion
exchanges between constituent quarks within different bags will not, in
general, be chiral symmetric. Thus the implementation of the
symmetry for models based on
non-relativistic quarks
requires families of diagrams which are more complex and involve
necessarily the binding potential.

In this work the quarks are bound by a scalar
field and the simplest chiral family of diagrams that encompasses
binding effects is related to the process $\pi q \rightarrow Sq$, which we
study in the sequence. Its amplitude is denoted by $ T_{\chi}$
and given by the diagrams displayed in fig.2. For {PV} coupling
there are just the direct (d) and crossed (x) diagrams, whereas for {PS}
coupling one has three possibilities, including a contact term.

In the {PV} coupling scheme, the amplitude $T_{\chi}$ is written as

\begin{equation}
T_\chi =-ig_s\frac g{2m}\tau _\alpha \bar{u}\left(\vec p^{\prime }\right)
\left[\frac{\rlap/p_d+m}{p_d^2-m^2}\rlap/k\gamma _5+ \rlap/k\gamma _5\frac{%
\rlap/p_x+m}{p_x^2-m^2}\right] u\left(\vec p\right) ,
\end{equation}
\label{36}

\noindent
with

\begin{eqnarray}
p_d &=&p+k, \\
p_x &=&p^{\prime }-k.  \nonumber
\end{eqnarray}
\label{37}

In order to show that this result is equivalent to that produced in the PS
approach, we use the Dirac equation and rewrite $T_{\chi}$ as

\begin{equation}
T_\chi =-ig_sg\tau _\alpha \bar{u}\left(\vec{p}^{\prime }\right) \left[\frac{%
\rlap/p_d+m}{p_d^2-m^2}\gamma _5+ \gamma _5\frac{\rlap/p_x+m}{p_x^2-m^2}%
+\frac 1m\gamma _5\right] u\left(\vec{p}\right) .
\end{equation}
\label{38}

This expression is the same one would obtain from the {PS} Lagrangian,
with the last term within the square brackets being due to the contact term
in fig.2. This result shows that the PV and PS schemes yield identical
results when the equations of motion for the external quarks can be used. On
the other hand, it also indicates that, as in the case of the OPEP, these
two approaches are not fully equivalent when one deals with off-shell
constituent quarks. Thus, in this case, the inclusion of 
first order effects 
in the scalar field improves the OPEP description, but does not
correspond to a complete solution of the problem. In fact, such a full
solution would require interactions in all orders of the potential.

Using the Dirac equation, one obtains a somewhat simpler form for
the amplitude

\begin{equation}
T_\chi =-ig_sg\tau _\alpha \bar{u}\left(\vec{p}^{\prime }\right) \left[ 
\frac{\rlap/k}{p_d^2-m^2}+\frac{\rlap/k}{p_x^2-m^2}+\frac 1m\right]
\gamma_5u\left(\vec{p}\right) .
\end{equation}
\label{39}

The $\pi N$ form factor is associated with the diagram shown in fig. 3A,
which involves an amplitude $T_\chi $ for each nucleon. However we cannot
use directly eq.(39) in the evaluation of the NN potential, for this would
produce an amplitude containing disconnected parts. In order to avoid
this, we must consider only the positive frequency irreducible part of
$ T_\chi $, which is shown in fig. 3B, and the proper NN interaction
is given by the diagrams (1-4) of fig. 3C.

For the one-body ($\pi q$) vertex in quark $v$ , we adopt the form given by
eq.(35), and have

\begin{equation}
\Gamma^{\left(v\right)}_{\pi q}=\left[g\tau _\alpha \gamma _5\right]^
{\left(v\right)} .
\end{equation}
\label{40}

The two body operator, on the other hand, is associated with the proper pion
diquark ($\pi d$) amplitude of fig.3B, which does not contain positive energy
intermediate states. In order to isolate these contributions,
we write the quark propagator as

\begin{equation}
\frac{\rlap/p+m}{p^2-m^2}=\frac 1{2E}\left[ \frac 1{p^0-E}\sum_su^s\left( 
\vec{p}\right) \bar{u}^s\left( \vec{p}\right) +\frac 1{p^0+E}\sum_sv^s\left(
-\vec{p}\right) \bar{v}^s\left( -\vec{p}\right) \right] ,
\end{equation}
\label{41}

\noindent
where

\begin{eqnarray}
E &=&\sqrt{\vec{p}^{\ 2}+m^2} .
\end{eqnarray}
\label{42}

Thus the contribution from the positive energy states is 

\begin{equation}
T_{\left( +\right) }=-ig_sg\tau _\alpha \bar{u}\left( \vec{p}^{\ \prime
}\right) \left[ \frac{\left( \rlap/\tilde{p}_d+m\right) }{2E_d\left(
p_d^0-E_d\right) }+\frac{\left( -\rlap/\tilde{p}_x+m\right) }{2E_x\left(
p_x^0-E_x\right) }\right] \gamma _5u\left( \vec{p}\right) ,
\end{equation}
\label{43}

\noindent with

\begin{eqnarray}
\tilde{p}_i &=&\left( E_i,\vec{p}_i\right) ,  \nonumber \\
E_i &=&\sqrt{\vec{p}_i^{\ 2}+m^2} ,
\end{eqnarray}
\label{44}

\noindent for $i = d, x$.

The diagrams of fig.3C involve scalar interactions between two quarks. If
one were dealing with just a perturbative exchange of a scalar particle of
mass $m_s$, as in the case of sigmas in nuclear physics, the evaluation of
this part of the diagram would yield a potential of the form

\begin{eqnarray}
\tilde{W_P}{\left(q_w\right)} &=& \frac{g_s^2}{{q_w}^2-m_s^2} ,
\end{eqnarray}
\label{45}

\noindent
where the subscript P stands for perturbation and $q_w$ is the exchanged
momentum,

\begin{eqnarray}
q_w &=& {p'_w}-p_w .
\end{eqnarray}
\label{46}

In the case of quarks, the scalar interaction, represented by a formal
momentum-space function $\tilde W( q_w )$, is highly non-linear and its
Fourier transform corresponds to the confining potential in configuration
space

\begin{eqnarray}
W {\left(r\right)} &=& \int \frac{d\vec{q_w}}{\left(2\pi \right) ^3} e^{-i%
\vec{q}_w\cdot\vec{r}}\tilde W\left( q_w \right) .
\end{eqnarray}
\label{47}

With this definition, the proper $\pi d$ amplitude for the quarks $v$ and $w$
of fig.3B is 

\begin{eqnarray}
T^{\left(vw\right)}_{\pi d} &=& -i\tilde W\left( q_w \right) \left[ g\tau
_\alpha \bar{u}\left(\vec{p}^{\ \prime }\right) \left( \frac{\rlap/k}{%
p_d^2-m^2}- \frac{\left(\rlap/\tilde{p}_d+m\right) }{2E_d\left(
p_d^0-E_d\right) }+ \frac{\rlap/k}{p_x^2-m^2} \right. \right.  \nonumber \\
&&\left. \left. - \frac{\left( -\rlap/\tilde{p}_x+m\right) }{%
2E_x\left(p_x^0-E_x\right) }+ \frac{1}{m}\right) \gamma _5u\left( \vec{p}%
\right) \right]^{(v)}\left[ \bar{u}\left( \vec{p}^{\ \prime }\right) u\left( 
\vec{p}\right) \right]^{(w)}.
\end{eqnarray}
\label{48}

\noindent
Using the Dirac equation, we have

\begin{eqnarray}
T^{\left(vw\right)}_{\pi d} &=&-i\tilde W\left( q_w \right) \left[ g\tau
_\alpha \bar{u}\left( \vec{p}^{\ \prime }\right) \left( \gamma ^0\frac{%
\left( E_x-E_d\right) }{2E_dE_x}+\frac 1m\right. \right.  \nonumber \\
&&\left. \left. -\rlap/k\left( \frac 1{2E_d\left( p_d^0+E_d\right) }+\frac
1{2E_x\left( p_x^0+E_x\right) }\right) \right) \gamma _5u\left( \vec{p}%
\right) \right] ^{\left( v \right) }\left[ \bar{u}\left( \vec{p}^{\ \prime
}\right) u\left( \vec{p}\right) \right] ^{\left( w\right) }.
\end{eqnarray}
\label{49}

Excluding the free spinors from this result, we obtain the proper $\pi d$
vertex as

\begin{eqnarray}
\Gamma^{\left(vw\right)} _{\pi d} &=&\tilde W\left( q_w \right) \left[ g\tau
_\alpha \left( \gamma ^0\frac{\left( E_x-E_d\right) }{2E_dE_x}+\frac
1m\right. \right.  \nonumber \\
&&\left. \left. -\rlap/k\left( \frac 1{2E_d\left( p_d^0+E_d\right) }+\frac
1{2E_x\left( p_x^0+E_x\right) }\right) \right) \gamma _5\right] ^{\left(
v\right) }I^{\left( w\right) } ,
\end{eqnarray}
\label{50}

\noindent
where $I^{\left( w\right) }$ is the identity matrix for the pure scalar
vertex in quark $w$ .


\section{Effective Potential}

In order to calculate the effective potential, we insert the $\pi q$ and $%
\pi d$ vertex functions between free spinors and use the following
correspondences between relativistic and non-relativistic amplitudes:

\begin{equation}
\bar{u}\left( \vec{p}_v^{\prime }\right) \Gamma^{\left(v\right)} _{\pi q}
u\left(\vec{p}_v\right) \stackrel{n.r.}{\rightarrow } - \frac{g}{2m}%
\tau_\alpha^{\left(v\right)} \vec{\sigma}^{\left(v\right)} \cdot \left( \vec{%
p}_v^{\prime} - \vec{p}_v\right) ,
\end{equation}
\label{51}

\begin{equation}
\bar{u}\left( \vec{p}_w^{\prime}\right) \bar{u}\left( \vec{p}_v^{\prime}
\right) \Gamma^{\left( vw\right)}_{\pi d} u\left( \vec{p}_v\right) u\left(%
\vec{p}_w\right) \stackrel{n.r.}{\rightarrow} \tilde W\left( q_w \right)
\left\{ {g\tau}_\alpha^{\left(v\right)} \left[ \frac{\vec{\sigma}
^{\left(v\right)}\cdot \vec{q_w}}{2m^2} \right] \right\} I^{\left(w\right)} .
\end{equation}
\label{52}

Using eqs.(51) and (52), we write the non-relativistic amplitudes
corresponding to the diagrams of fig.3C, in the centre of mass (CM) frame of the
NN system, as

\begin{eqnarray}
t_{qq}^{\left(v,y\right)} &=& \left( \frac{g}{2m}\right)^2 \vec{\tau}%
^{\left(v\right)} \cdot \vec{\tau}^{\left(y\right)} \left[ \vec{\sigma}%
^{\left(v\right)} \cdot \vec{k}\right] \frac{1}{\vec{k}^2+ m_\pi ^2} \left[ 
\vec{\sigma}^{\left(y\right)}\cdot \vec{k} \right] , \\
t_{dq}^{\left(vw,y\right)} &=& -\left( \frac{g}{2m}\right)^2 \vec{\tau}%
^{\left(v\right)}\cdot \vec{\tau}^{\left(y\right)} \left[ \vec{\sigma}%
^{\left(v\right)}\cdot \vec{q}_w {\left(\frac{\tilde W\left( q_w \right)}{m}%
\right)} I^{\left(w\right)}\right] \frac{1}{\vec{k}^2 + m_\pi^2} \left[ \vec{%
\sigma}^{\left(y\right)}\cdot \vec{k}\right] , \\
t_{qd}^{\left(v,yz\right)} &=&\left( \frac{g}{2m}\right)^2 \vec{\tau}%
^{\left(v\right)}\cdot \vec{\tau}^{\left(y\right)} \left[\vec{\sigma}%
^{\left(v\right)}\cdot \vec{k}\right] \frac {1}{\vec{k}^2 + m_\pi ^2} \left[%
\vec{\sigma}^{\left(y\right)}\cdot \vec{q}_z {\left(\frac{\tilde W\left( q_z
\right)}{m}\right)}I^{\left(z\right)}\right] , \\
t_{dd}^{\left(vw,yz\right)} &=&-\left( \frac{g}{2m}\right)^2 \vec{\tau}%
^{\left(v\right)}\cdot \vec{\tau}^{\left(y\right)} \left[ \vec{\sigma}%
^{\left(v\right)}\cdot \vec{q}_w {\left(\frac{\tilde W\left( q_w \right)}{m}%
\right)}I^{\left(w\right)}\right]  \nonumber \\
& \times & \frac{1}{\vec{k}^2 + m_{\pi}^2} \left[ \vec{\sigma}%
^{\left(y\right)}\cdot \vec{q}_z {\left(\frac{\tilde W\left( q_z \right)}{m}%
\right)} I^{\left(z\right)}\right].
\end{eqnarray}
\label{56}

These results are related to the potentials between quarks and diquarks in
momentum space by

\begin{equation}
< \vec{p'}_1...\vec{p'}_6 | V | \vec{p}_1...\vec{p}_6 > = 
- (2\pi)^3 \delta 
{\left[(\vec{p'}_1 +...+ \vec{p'}_6) 
- (\vec{p}_1+...+\vec{p}_6)\right]} \; t .
\end{equation}
\label{57}

The Fourier transform of this expression yields a potential in coordinate
space, that is local and given by

\begin{equation}
<\vec{r'}_1...\vec{r'}_6 | V | \vec{r}_1...\vec{r}_6 > =
\delta (\vec{r'}_1 - \vec{r}_1)...
\delta (\vec{r'}_6 - \vec{r}_6) 
\;\Pi \;(\vec{r}_1...\vec{r}_6),
\end{equation}
\label{58}

\noindent
where $\Pi$ is the potential due to pion exchanges, as defined in
section 2. Using 

\begin{equation}
\vec r_{ij} = \vec r_i - \vec r_j 
\end{equation}
\label{59}

\noindent
and results (53-56), we obtain the potentials in configuration 
space

\begin{eqnarray}
   \Pi ^{\left(v,y\right)}_{qq} &=&-\left( \frac{g}{2m}\right)^2 \vec{\tau}%
^{\left(v\right)}\cdot\vec{\tau}^{\left(y\right)} {\sigma}^{\left(v\right)}_i%
{\sigma}^{\left(y\right)}_j \left[\int \frac{d\vec{k}}{\left( 2\pi\right) ^3}
\frac{k^ik^j}{\vec{k}^2+m_\pi ^2} e^{-i\vec{k}\cdot\vec{r}_{vy}}\right], \\
   \Pi^{\left(vw,y\right)}_{dq} &=&\left( \frac{g}{2m}\right)^2 \vec{\tau}%
^{\left(v\right)}\cdot\vec{\tau}^{\left(y\right)} {\sigma}%
^{\left(v\right)}_i {\sigma}^{\left(y\right)}_j \left[\int \frac{d\vec{q}_w}{%
\left(2\pi \right)^3} {q}_w^i{\left(\frac{\tilde W\left( q_w \right)}{m}%
\right) e^{-i\vec{q}_w\cdot\vec{r}_{wv}}}\right]  \nonumber \\
&\times&\left[\int \frac{d\vec{k}}{\left( 2\pi \right)^3} \frac{k^j}{\vec{k}%
^2+m_\pi ^2} e^{-i\vec{k}\cdot\vec{r}_{vy}}\right], \\
   \Pi^{\left(v,yz\right)}_{qd} &=&-\left( \frac{g}{2m}\right)^2 \vec{\tau}%
^{\left(v\right)}\cdot\vec{\tau}^{\left(y\right)} {\sigma}%
^{\left(v\right)}_i {\sigma}^{\left(y\right)}_j \left[\int \frac{d\vec{k}}{%
\left( 2\pi\right)^3} \frac{k^i}{\vec{k}^2+m_\pi^2} e^{-i\vec{k}\cdot\vec{r}%
_{vy}}\right]  \nonumber \\
&\times& \left[\int \frac{d\vec{q}_z}{\left( 2\pi \right) ^3} {q}^j_z{\left(%
\frac{\tilde W\left( q_z \right)}{m}\right)} e^{-i\vec{q}_z\cdot\vec{r}%
_{zy}}\right], \\
   \Pi^{\left(vw,yz\right)}_{dd} &=&\left(\frac{g}{2m}\right)^2 \vec{\tau}%
^{\left(v\right)}\cdot\vec{\tau}^{\left(y\right)} {\sigma}%
^{\left(v\right)}_i {\sigma}^{\left(y\right)}_j \left[\int \frac{d\vec{q}_w}{%
\left( 2\pi \right)^3} {q}_w^i{\left(\frac{\tilde W\left( q_w \right)}{m}%
\right)} e^{-i\vec{q}_w\cdot\vec{r}_{wv}}\right]  \nonumber \\
&\times& \left[\int\frac{d\vec{k}}{\left( 2\pi \right) ^3} \frac{1}{\vec{k}%
^2+m_\pi^2} e^{-i\vec{k}\cdot\vec{r}_{vy}}\right] \left[\int\frac{d\vec{q}_y%
}{\left( 2\pi \right)^3} {q}_z^j {\left(\frac{\tilde W\left( q_z \right)}{m}%
\right)} e^{-i\vec{q}_z\cdot\vec{r}_{zy}}\right] .
\end{eqnarray}
\label{60,63}

The full potential due to pion exchanges is thus given by

\begin{eqnarray}
\Pi _{ab}^\chi &=& \sum_{v}\sum_{y}\Pi^{\left(v,y\right)}_{qq}
+\sum_{v,w}\sum_{y}\Pi^{\left(vw,y\right)}_{dq}
+\sum_{v}\sum_{y,z}\Pi^{\left(v,yz\right)}_{qd}  \nonumber \\
&+& \sum_{v,w}\sum_{y,z}\Pi^{\left(vw,yz\right)}_{dd} ,
\end{eqnarray}
\label{64}

\noindent
where the symbol $\sum_{i,j}$ indicates a sum over $i$ and $j$, with 
$i\neq j$.

This is the pion exchange potential between quarks in
different nucleons. In order to obtain the effective potential, we use this
result in eq.(17) and have

\begin{eqnarray}
&&V_{NN}\left( \vec{R}_a,\vec{R}_b\right) = \left( \frac{g}{2m}\right)^2
\int d\Omega \left\langle I_b \right|\left\langle I_a \right|  \nonumber \\
&\times& \left\{ \sum_{v}\sum_{y} \vec{\tau}^{\left(v\right)}\cdot \vec{\tau}%
^{\left(y\right)} \left[{\vec{\sigma}^{\left(v\right)}\cdot\vec{\nabla} \vec{%
\sigma}^{\left(y\right)}\cdot \vec{\nabla} U\left( \vec{r}_{vy}\right)}%
\right] \right.  \nonumber \\
&-& \sum_{v,w}\sum_{y} \vec{\tau}^{\left(v\right)}\cdot \vec{\tau}%
^{\left(y\right)} \left[\vec{\sigma}^{(v)}\cdot\vec{\nabla} {\left(\frac{%
W\left(\vec{r}_{wv}\right)}{m}\right)}\right] \left[\vec{\sigma}^{(y)}\cdot%
\vec{\nabla}U\left(\vec{r}_{vy}\right)\right]  \nonumber \\
&+&\sum_{v}\sum_{y,z} \vec{\tau}^{\left(v\right)}\cdot \vec{\tau}%
^{\left(y\right)} \left[ \vec{\sigma}^{(v)}\cdot \vec{\nabla}U\left( \vec{r}%
_{vy}\right) \right] \left[\vec{\sigma}^{(y)}\cdot\vec{\nabla} {\left(\frac{%
W\left(\vec{r}_{zy}\right)}{m}\right)}\right] \\
&-&\left.\sum_{v,w}\sum_{y,z} \vec{\tau}^{\left(v\right)}\cdot \vec{\tau}%
^{\left(y\right)} \left[\vec{\sigma}^{(v)}\cdot\vec{\nabla} {\left(\frac{%
W\left(\vec{r}_{wv}\right)}{m}\right)}\right] U \left( \vec{r}_{vy}\right)
\left[\vec{\sigma}^{(y)}\cdot\vec{\nabla} {\left(\frac{W\left(\vec{r}%
_{zy}\right)}{m}\right)}\right] \right\} \left| I_a\right\rangle
\left|I_b\right\rangle,  \nonumber
\end{eqnarray}
\label{65}

\noindent
where $W(\vec r)$ is the configuration space confining potential and $U(\vec
r)$ is the Yukawa function

\begin{eqnarray}
U{(\vec r)}&=&\frac{4\pi}{\mu} \int \frac{d\vec{k}}{\left( 2\pi \right) ^3} 
\frac{e^{-i\vec{k}\cdot\vec{r}}}{{k}^2+\mu^2}  \nonumber \\
&=&\frac{e^{{-\mu}r}}{{\mu}r}.
\end{eqnarray}
\label{66}

This is the main result of this work. The first term within the curly
brackets is the usual OPEP between quarks whereas the other ones correspond
to corrections due to chiral symmetry, in the form of gradients of the
confining potential. An interesting feature of this result is that all the
terms of the effective potential contain two gradients, reflecting the fact
that they come from a uniform expansion in momentum space, as expected from
a calculation based on chiral symmetry.


\section{Application}

In this section we apply the results from the previous section to the case
of a nucleon composed by three quarks bound by a harmonic potential of the
form

\begin{eqnarray}
W\left(\vec r\right) = \frac {1}{2} K r^2 .
\end{eqnarray}
\label{67}

\noindent
In order to make the structure of our calculation more transparent, we allow
different confining constants for the two nucleons.

The internal nucleon wave function is given by

\begin{eqnarray}
\left| I\right\rangle &=& \left| \vec{\rho},\vec{\lambda},S^z,T^z,C\right%
\rangle  \nonumber \\
&=& \frac{\alpha ^3}{\pi ^{3/2}} e^{-\frac{\alpha^2}{2}\left(\vec{\rho}^2+%
\vec{\lambda}^2\right)} \left|S^z\right\rangle\left|T^z\right\rangle \left|
C\right\rangle
\end{eqnarray}
\label{68}

\noindent
where $\vec \rho $ and $\vec \lambda $ are Jacobi coordinates and $S^z$, $I^z
$ and C are spin, isospin and color states. The
color component $\left| C\right\rangle $ is totally antisymmetric with
respect to quark permutations and the same happens with the full
wave-function. The constant $\alpha $ represents the size of the nucleon, is
given by $\alpha ^2=\sqrt{3Km}$ and related to the binding energy per
nucleon by $\omega ={\frac{\alpha ^2}m}$.

The action of the quark spin and isospin operators over the nucleon wave
function is related to the corresponding collective operators by

\begin{eqnarray}
\tau ^{(v)}_i\left| S^z,I^z\right\rangle &\equiv& \frac{1}{3}
{\cal{T}}_i\left|S^z,I^z\right\rangle +...\;\;, \\
\sigma ^{(v)}_i\left| S^z,I^z\right\rangle &\equiv& \frac{1}{3}\Sigma
_i\left|S^z,I^z\right\rangle +...\;\;, \\
\sigma ^{(v)}_i\tau ^{(v)}_j\left| S^z,I^z\right\rangle &\equiv& \frac{5}{9}%
\Sigma_i {\cal{T}}_j \left| S^z,I^z\right\rangle +...\;\;, \\
\sigma ^{(v)}_i\tau ^{(w)}_j\left| S^z,I^z\right\rangle &\equiv& -\frac{1}{9}
\Sigma_i {\cal{T}}_j\left| S^z,I^z\right\rangle +...v\neq w\;\;,
\end{eqnarray}
\label{69,72}

\noindent
where we have omitted non-nucleon states on the right hand side. Using these
results in eqs.(60-64), we obtain the effective potential operator in spin
and isospin spaces

\begin{eqnarray}
&&V_{NN}\left( \vec{R}_a,\vec{R}_b\right) = -\left( \frac{g}{2m}\frac
59\right)^2 \frac{\alpha _a^6\alpha _b^6}{\pi ^3\pi ^3}\; 
{\cal{\vec{T}}}^{(a)}\cdot{\cal{\vec{T}}}^{(b)} 
\Sigma_i^{(a)}\Sigma _j^{(b)} 
\nonumber \\
&\times&\left\{ \sum_{v}\sum_{w} \int \frac{d\vec{k}}{\left( 2\pi \right) ^3}
\frac{k^ik^j}{{k}^2+m_\pi^2} \int d\Omega e^{-\alpha_a^2\left( \vec{\rho}%
_a^2+\vec{\lambda}_a^2\right) -\alpha _b^2\left( \vec{\rho}_b^2+\vec{\lambda}%
_b^2\right) -i\vec{k}\cdot\vec{r}_{vy}} \right.  \nonumber \\
&-& i\frac{K_a}{m} \sum_{v,w}\sum_{y} \int \frac{d\vec{k}}{\left( 2\pi
\right) ^3} \frac{k^j}{{k}^2+m_\pi ^2} \int d\Omega{r}_{wv}^i e^{-\alpha
_a^2\left( \vec{\rho}_a^2+\vec{\lambda}_a^2\right) -\alpha_b^2\left( \vec{%
\rho}_b^2+\vec{\lambda}_b^2\right) -i\vec{k}\cdot\vec{r}_{vy}}  \nonumber \\
&+&i\frac{K_b}{m} \sum_{v}\sum_{y,z} \int \frac{d\vec{k}}{\left(2\pi \right)
^3} \frac{k^i}{{k}^2+m_\pi ^2} \int d\Omega{r}_{zy}^j e^{-\alpha
_a^2\left( \vec{\rho}_a^2+\vec{\lambda}_a^2\right) -\alpha_b^2\left( \vec{%
\rho}_b^2+\vec{\lambda}_b^2\right) -i\vec{k}\cdot\vec{r}_{vy}}  \nonumber \\
&-&\left . i^2\frac{K_aK_b}{m^2} \sum_{v,w}\sum_{y,z} \int \frac{d\vec{k}}{%
\left( 2\pi \right) ^3}\frac 1{{k}^2+m_\pi ^2} \int d\Omega{r}_{wv}^i {r}%
_{zy}^j e^{-\alpha _a^2\left( \vec{\rho}_a^2+\vec{\lambda}_a^2\right)
-\alpha_b^2\left( \vec{\rho}_b^2+\vec{\lambda}_b^2\right) -i\vec{k}\cdot\vec{%
r}_{vy}} \right\}  \nonumber \\
&\times &\left| S_a^z,I_a^z,S_b^z,I_b^z\right\rangle .
\end{eqnarray}
\label{73}

The vector $\vec r_{vy}$ that enters these expressions may be written as
linear combinations of $\vec R, \vec\rho$ and $\vec\lambda$

\begin{eqnarray}
\vec r_{vy} = \vec{X}+ \left(c_{v\rho}\vec{\rho}_a+c_{v\lambda}\vec{\lambda}%
_a -c_{y\rho }\vec{\rho}_b-c_{y\lambda }\vec{\lambda}_b \right) ,
\end{eqnarray}
\label{74}

\noindent
where the coefficients $c_{ij}$ have the values $c_{1\rho}={\sqrt\frac{1}{2}}
$, $c_{2\rho}=-{\sqrt\frac{1}{2}}$, $c_{3\rho}=0$, $c_{1\lambda}={\sqrt\frac{%
1}{6}}$, $c_{2\lambda}={\sqrt\frac{1}{6}}$, $c_{3\lambda}=-{\sqrt\frac{2}{3}}
$ and obey the relationships

\begin{eqnarray}
&&c_{i\rho}^2+c_{i\lambda}^2=\frac{2}{3}, \\
&&\left(c_{i\rho}-c_{j\rho}\right)c_{i\rho}+
\left(c_{i\lambda}-c_{j\lambda}\right)c_{i\lambda}=1 .
\end{eqnarray}
\label{75,76}

These results allow the various gaussian integrations to be performed and we
obtain

\begin{eqnarray}
I_{(v,y)} &=&\frac{\alpha _a^6\alpha _b^6}{\pi ^3\pi ^3}\int d\Omega
e^{-\alpha_a^2\left(\vec{\rho}_a^2+\vec{\lambda}_a^2\right)
-\alpha_b^2\left( \vec{\rho}_b^2+\vec{\lambda}_b^2\right) -i\vec{k}%
\cdot\left(c_{v\rho }\vec{\rho}_a+c_{v\lambda }\vec{\lambda}_a -c_{w\rho }%
\vec{\rho}_b-c_{w\lambda }\vec{\lambda}_b\right)}  \nonumber \\
&=&e^{-A{k}^2}, \\
I_{(wv,y)}^i &=&\frac{\alpha _a^6\alpha _b^6}{\pi ^3\pi ^3}\int d\Omega
e^{-\alpha_a^2\left(\vec{\rho}_a^2+\vec{\lambda}_a^2\right) -\alpha
_b^2\left(\vec{\rho}_b^2+\vec{\lambda}_b^2\right) -i\vec{k}\cdot \left(
c_{v\rho }\vec{\rho}_a+c_{v\lambda }\vec{\lambda}_a -c_{y\rho }\vec{\rho}%
_b-c_{y\lambda }\vec{\lambda}_b\right)}  \nonumber \\
&\times& \left[\left(c_{w\rho}-c_{v\rho}\right)\vec{\rho}_a+
\left(c_{w\lambda}-c_{v\lambda }\right) \vec{\lambda}_a\right]^i  \nonumber
\\
&=&i{\frac{k^{i}}{2\alpha_a^2}}e^{-A{k}^2}, \\
I_{(v,zy)}^j &=&\frac{\alpha _a^6\alpha _b^6}{\pi ^3\pi ^3}\int d\Omega
e^{-\alpha_a^2\left(\vec{\rho}_a^2+\vec{\lambda}_a^2\right) -\alpha
_b^2\left( \vec{\rho}_b^2+\vec{\lambda}_b^2\right) -i\vec{k}\cdot \left(
+c_{v\rho }\vec{\rho}_a+c_{v\lambda }\vec{\lambda}_a -c_{y\rho }\vec{\rho}%
_b-c_{y\lambda }\vec{\lambda}_b\right)}  \nonumber \\
&\times& \left[\left( c_{z\rho }-c_{y\rho }\right) \vec{\rho}_b
+\left(c_{z\lambda }-c_{y\lambda }\right) \vec{\lambda}_b\right] ^j 
\nonumber \\
&=&-i{\frac{k^{j}}{2\alpha_b^2}}e^{-A{k}^2}, \\
I_{(wv,zy)}^{ij} &=&\frac{\alpha _a^6\alpha _b^6}{\pi ^3\pi ^3}\int d\Omega
e^{-\alpha _a^2\left(\vec{\rho}_a^2+\vec{\lambda}_a^2\right) -\alpha
_b^2\left( \vec{\rho}_b^2+\vec{\lambda}_b^2\right) -i\vec{k}\cdot\left(
c_{v\rho }\vec{\rho}_a+c_{v\lambda }\vec{\lambda}_a -c_{y\rho }\vec{\rho}%
_b-c_{y\lambda }\vec{\lambda}_b\right) }  \nonumber \\
&\times&\left[\left( c_{w\rho }-c_{v\rho }\right)\vec{\rho}_a +\left(
c_{w\lambda }-c_{v\lambda }\right) \vec{\lambda}_a\right]^i \left[\left(
c_{z\rho }-c_{y\rho }\right) \vec{\rho}_b +\left( c_{z\lambda}-c_{y\lambda
}\right) \vec{\lambda}_b\right]^j  \nonumber \\
&=&{\frac{k^{i}}{2\alpha_a^2}}{{\frac{k^{j}}{2\alpha_b^2}}}e^{-A{k}^2},
\end{eqnarray}
\label{77,80}

\noindent
where

\begin{eqnarray}
A=\left(\frac1{6\alpha _a^2}+\frac 1{6\alpha _b^2}\right).
\end{eqnarray}
\label{81}

Thus all configuration space integrals entering eq.(73) become proportional
to

\begin{eqnarray}
U^{ij}\left(\vec X,\alpha_a,\alpha_b\right) &=&\frac{4\pi}{\mu} \int \frac{d%
\vec{k}}{\left( 2\pi \right) ^3} \frac{k^ik^j}{{k}^2+\mu^2} 
e^{\left(-i\vec{k}\cdot\vec X -A{k}^2\right)}.
\end{eqnarray}
\label{82}

Using these results in eq.(73), we obtain

\begin{eqnarray}
V_{NN}\left(\vec X,\alpha_a,\alpha_b\right) &=& \left( \frac{g}{2m}\frac
53\right)^2 \left( 1+\frac{\alpha _a^2}{3m^2}\right) \left( 1+\frac{\alpha
_b^2}{3m^2}\right) \frac{\mu}{4\pi}  \nonumber \\
&\times &{\cal{\vec{T}}}^{(a)}\cdot {\cal{\vec{T}}}^{(b)} \vec{\Sigma%
}^{(a)}\cdot\vec{\nabla} \vec{\Sigma}^{(b)}\cdot \vec{\nabla}\ U\left(\vec
X,\alpha_a,\alpha_b\right) .
\end{eqnarray}
\label{83}

\noindent where

\begin{eqnarray}
U\left(\vec X,\alpha_a,\alpha_b\right) &=& \frac{4\pi }{\mu}\int \frac{d\vec{%
k}}{\left( 2\pi\right)^3} 
\frac{e^{\left(-i\vec{k}\cdot\vec{X}-Ak^2\right)}}
{\vec{k}^2+\mu^2}
\nonumber \\
&=&\frac 2{\pi \mu}\int dk \frac{k^2e^{-Ak^2}}{{k}^2+\mu^2}%
j_0\left(kX\right) .
\end{eqnarray}
\label{84}

This integral may be performed analytically and we have

\begin{eqnarray}
U\left(\vec X,\alpha_a,\alpha_b\right) &=& -\frac{e^{A\mu^2}}{\mu X} 
\left[ \sinh {\mu X}
- \frac{1}{2}e^{-\mu X}erf\left(\frac{X}{2\sqrt{A}}-\mu\sqrt{A}\right)
\right.  
\nonumber \\
&-&\left. 
\frac{1}{2}e^{\mu X}erf\left(\frac X{2\sqrt{A}}+\mu\sqrt{A}\right)\right] .
\end{eqnarray}
\label{85}

\noindent
For large values of $X$ this integral becomes

\begin{equation}
U\left(\vec X,\alpha_a,\alpha_b\right) \stackrel{x\rightarrow \infty }{%
\Rightarrow } e^{\frac{\mu^2}{6\alpha _a^2}} e^{\frac{\mu^2}{6\alpha _b^2}}\; 
\frac{e^{-\mu X}}{\mu X} ,
\end{equation}
\label{86}

\noindent
after using $erf(\infty) = 1$.

The potential therefore reduces to

\begin{eqnarray}
V_{NN}\left( \vec{X}\right) &=&\left( \frac{g}{2m}\frac 53\right)^2 \frac{\mu%
}{4\pi } \left( 1+\frac{\alpha _a^2}{3m^2}\right)e^{\frac{\mu^2}{6\alpha _a^2%
}} \left( 1+\frac{\alpha _b^2}{3m^2}\right)e^{\frac{\mu^2}{6\alpha _b^2}} 
\nonumber \\
&\times &{\cal{\vec{T}}}^{(a)}\cdot {\cal{\vec{T}}}^{(b)} \; \vec{\Sigma%
}^{(a)}\cdot \vec{\nabla} \; \vec{\Sigma}^{(b)}\cdot \vec{\nabla} \; \left(\frac{%
e^{-\mu X}}{\mu X}\right) .
\end{eqnarray}
\label{87}

Comparing this result with the usual expression for the OPEP, we have

\begin{equation}
\frac 53\frac{g}{2m}\exp\left(\frac{\mu^2}{6\alpha^2}\right) \left(1+\frac{%
\alpha^2}{3m^2}\right) = \frac{g_{\pi NN}}{2M_N}.
\end{equation}
\label{88}

Going back to eq.(83), we write

\begin{equation}
V_{NN}\left( \vec{X}\right) =\left( \frac {g_{\pi NN}}{2M}\right)^2 {%
\cal{\vec{T}}}^{(a)}\cdot {\cal{\vec{T}}}^{(b)} \; \vec{\Sigma}%
^{(a)}\cdot \vec{\nabla}\; \vec{\Sigma}^{(b)}\cdot \vec{\nabla} \int \frac{d%
\vec{k}}{\left( 2\pi \right) ^3}\frac{e^{-i \vec{k}\cdot \vec{X}-\left({%
k}^2+\mu^2\right) A}}{\vec{k} ^2+\mu^2} .
\end{equation}
\label{89}

This expression allows the $\pi N$ form factor to be identified as

\begin{equation}
\bar{G}({k})=\exp \left( -\frac{\vec{k}^2+\mu^2}{6\alpha^2} \right).
\end{equation}
\label{90}

Thus one learns that, in the case of a harmonic confining
potential, the $\pi N$ form factor is not modified by the 
binding corrections, since it is the same as one would
obtain by considering only the diagram 1 of fig.3C.

On the other hand, eq.(88) allows one to write the effective pion-quark
coupling constant as

\begin{equation}
g_{eff}= g\left(1+\frac{\omega}{3m}\right),
\end{equation}
\label{91}

\noindent
with $\omega =\sqrt{\frac{3K}m}$, indicating that it is
influenced by binding corrections. In order to interpret this result, we go
back to the PS Lagrangian given by
eq.(33) and note that, within the approximations considered here, it
is equivalent to having the field S replaced by a mean value such that

\begin{equation}
{\frac{g_s}{f_\pi}}\langle S \rangle = g\frac{\omega}{3m}
\end{equation}
\label{92}

\noindent
and this corresponds to the effective mass

\begin{equation}
m_{eff}= g f_\pi\left(1+\frac{\omega}{3m}\right).
\end{equation}
\label{93}

\noindent
In the case of the PV Lagrangian, on the other hand, eq.(92)
produces a shift in the quark mass which translates into a change of
the coupling constant when the effective equation of motion is 
used in the $ \pi$N vertex.

The shift in the effective mass
is due to the potential energy associated with each particle,
given by the
expectation value of the function $\frac{K}{2}r_i^2$ (and not $\frac{K}{2}
r_{ij}^2 !$), which yields $\frac{\omega}{3}$. Alternatively, one may note
that the total potential energy in the nucleon is $\frac{3}{2}\omega$; of
this amount, $\frac{1}{2}\omega$ corresponds to the energy of the CM
system and hence $\frac{\omega}{3}$ is available for shifting each quark
mass.


\section{Conclusions}

In this work we have studied the role of chiral symmetry in the $\pi N$ form
factor, assuming the nucleon to be a constituent quark cluster, bound by
a scalar field. The symmetry was implemented by means of two
different Lagrangians, for $PS$ and $PV$ pion quark couplings, which led to
identical results, as expected in a consistent symmetrical calculation.

The implementation of chiral symmetry in a system of bound quarks requires
the use of both one and two quark operators, the latter corresponding
to binding corrections. This gives rise to a nucleon-nucleon effective
potential that is a uniform second order polynomial in the pionic and
internal momenta. In configuration space, it contains two gradients, acting
on either the Yukawa function or on the confining potential.

In order to assess the properties of this chiral effective potential in a
particular model, we considered the case of a harmonic confining force
characterized by a frequency $\omega $. In this instance, the
shape of the $\pi N$ form factor is not modified by the symmetry, since the
harmonic wave function is an exponential and hence does not change its form
upon derivation. On the other hand, the $\pi N$ coupling constant receives
chiral contributions of the order $\frac \omega m$, $m$ being the
constituent quark mass. The fact that in many models we find $\omega \sim m$
means that these corrections are significant and tend to favour smaller
values of the coupling constant. In some baryon models, pion exchanges are
used to generate spin dependent forces and hence changes in the coupling
constant may influence the observables.

It is important to stress that our calculation deals only with the internal
part of the form factor and, in particular, effects associated with the pion
cloud were not considered. Therefore the phenomenological implications of our
study cannot be fully explored at present, but the consistent inclusion of
cloud effects is part of our programme.

However, even with its present limitations, our work provides an insight on
how the symmetry works in bound systems. It is a general feature of chiral
models that fermion masses and coupling constants to pions be constrained by
the Goldberger-Treiman relation. Therefore one expects that any chiral
interaction which modifies the coupling constant should also shift the
constituent quark mass. In the case of the harmonic potential, we have shown
that this coherent picture emerges from the inclusion of binding effects.


\newpage
\vspace{4cm} {Acknowledgments} \vspace{1cm}

It is our pleasure to thank Dr. Gast\~ao I. Krein for drawing our attention
to this problem and for numerous discussions. The work of one of us (C.M.M.)
was supported by FAPESP, Brazilian Agency.


\newpage
{\large \bf{Figure Captions}} \vspace{2cm}

\noindent
Fig.1. Pion cloud contribution to the $\pi N$ form factor; pions and
nucleons are represented by dashed and continuous lines respectively.

\noindent
Fig.2. Chiral amplitude for the process $\pi q\rightarrow S q$, where the
scalar boson is represented by a wavy line and (PV) and (PS)
stand for pseudovector and
pseudoscalar couplings.

\noindent
Fig.3. A: interaction between two clusters; 
B: one quark irreducible pion-diquark vertex, where
the propagator with the insertion (+) corresponds to positive
frequency states;
C: the proper part of the NN interaction.
\newpage
\begin{figure}
\epsfbox{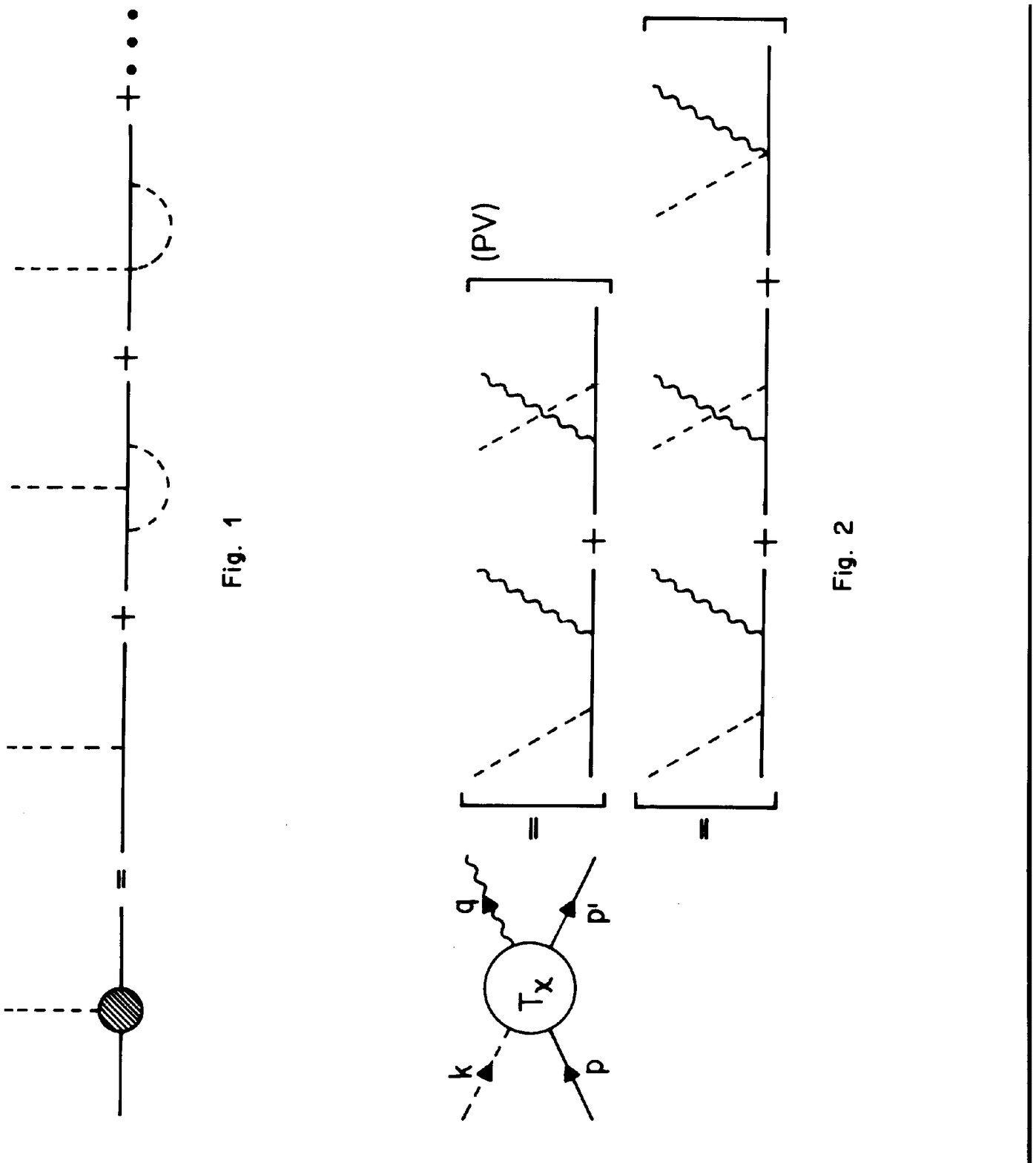}
\end{figure}
\newpage
\begin{figure}
\epsfbox{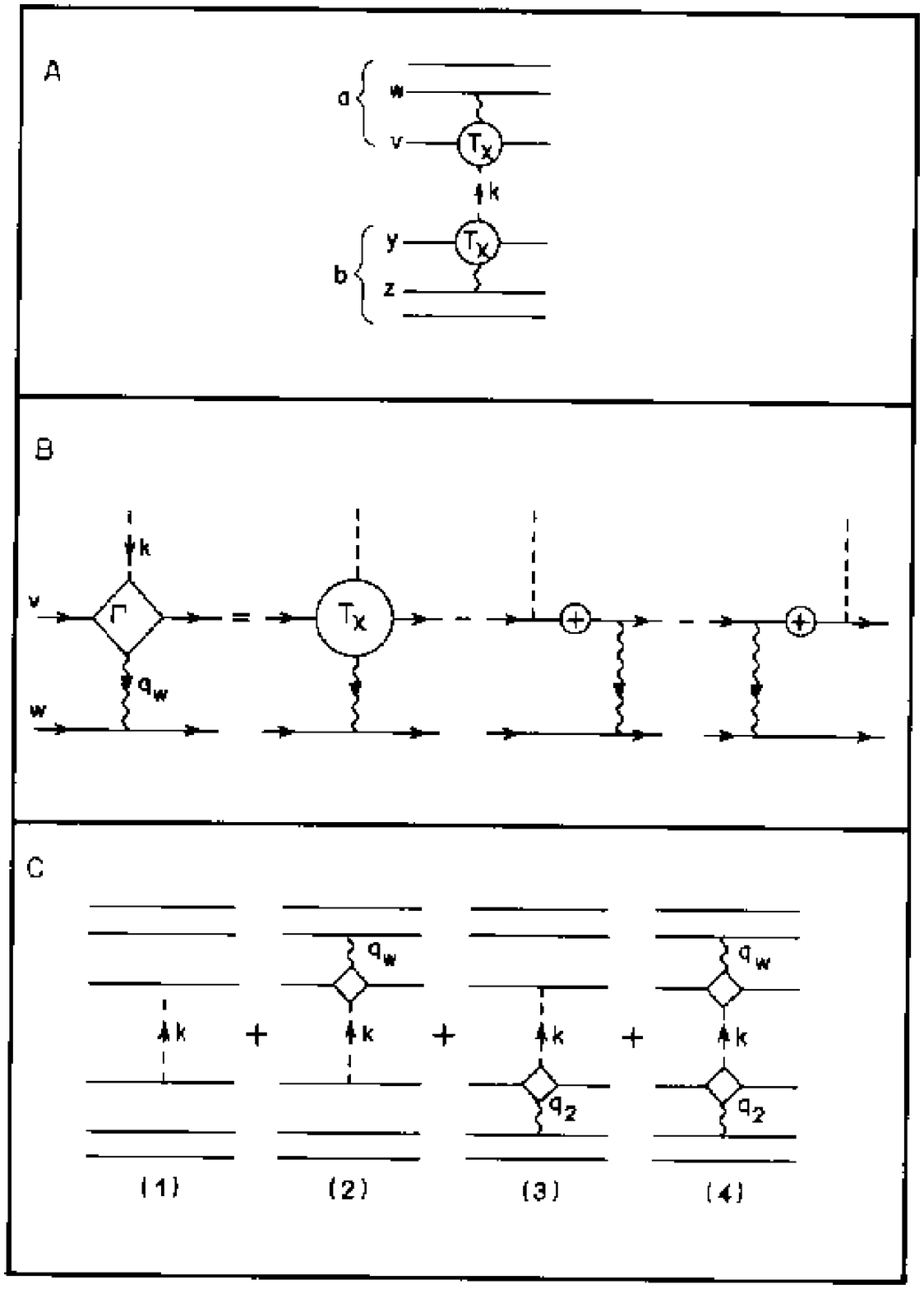}
\end{figure}


\begin{thebibliography}{99}
\bibitem{ref1}  for a review, see V. Bernard, N. Kaiser and Ulf-G. Meissner,
Int. J. Mod. Phys. E4,193(1995).

\bibitem{ref2}  A. W. Thomas, Adv. in Nucl. Phys.13,1(1990); G. A. Miller,
Int. Rev. of Nucl. Phys., ed. by W. Weise,1(1984).

\bibitem{ref3}  L. Wilets, $Non-TopologicalSolitons$, World Scientific,
Singapore (1989)

\bibitem{ref4}  J. F. Mathiot, Nucl. Phys. A412,201(1984); Nucl. Phys. A446,
123c(1985).

\bibitem{ref5}  O. L. Battistel and M. R. Robilotta, Phys. Rev.
C48,920(1993).

\bibitem{ref6}  R. Friedberg and T. D. Lee, Phys. Rev.D15,1694(1977).

\bibitem{ref7}  A. G. Williams and L. R. Dodd, Phys. Rev. D37,1971(1988).

\bibitem{ref8}  G. Krein, P. Tang, L. Wilets and A.G. Willians, Phys. Lett.
B212,362(1988); Nucl. Phys. A523,548(1991).

\bibitem{ref9}  S. Weinberg, Phys. Rev. Lett. 18,188(1967).

\bibitem{ref10}  S. Weinberg, Phys. Rev. 166,1568(1968).

\bibitem{ref11}  S. Coleman, J. Wess and B. Zumino, Phys. Rev.
177,2239(1969); C. G. Callan,S. Coleman, J. Wess and B. Zumino, Phys. Rev.
177,2247(1969).

\bibitem{ref12}  S. A. Coon and J. L. Friar, Phys. Rev. C 34,1060(1986).

\bibitem{ref13}  C. Ordo\~nez and U. Van Kolk, Phys. Lett. B291, 459 (1992).
\end{thebibliography}
\end{document}